\documentclass[aps,preprint,showpacs,showkeys]{revtex4}
\usepackage{graphicx}
\usepackage{amssymb}
\usepackage{bm}
\begin{document}
\setcounter{page}{1}
\title[]{Effects of the $\gamma-$rays Scattered Backward by Metals \\
on the Nuclear Energy Level Width
 }
\author{Il-Tong \surname{Cheon}}
\email{itcheon@gmail.com} \affiliation{The Korean Academy of
Science and Technology, 7-1 Gumi-dong, Bundang-gu, Seongnam-shi,
Gyunggi-do 463-808, Korea}
\author{Moon Taeg \surname{Jeong}}
\email{mtjeong@naver.com}
\thanks{Fax: +82-61-330-3309}
\affiliation{Department of Radiological Science, Dongshin
University, Naju 520-714, Korea}

\begin{abstract}
By placing a ${}^{133}Cs$ $\gamma$-ray source embedded in a solid
at the center of a platinum (gold) cylinder, we try to change the
width of the $81$-keV level. Our results show a narrowed energy
level and, equivalently, a prolonged lifetime. With a
0.5-mm-thick, 5-cm-long, 2-mm-diameter platinum cylinder, we
obtain a width narrower by $6.1 \% $ at $4.2 \: K$.
\end{abstract}



\maketitle
\section{Introduction}
Recently, the $\gamma$-ray backward scattering cross sections were
measured and compared with the theoretical predictions \cite{01,
02}. The results show that some of the $\gamma$-rays can be
scattered completely backward in a coherent and elastic way. This
fact implies that a $\gamma$-ray emitted from a radioactive
nucleus can return to the source nucleus without any energy loss
when a suitable reflector is applied.

The nuclear energy level width has been known as a solidly
determined and unchangeable quantity. Although there have been
several attempts to change it by altering the chemical state
\cite{03, 04, 05} and by applying high pressure \cite{06} or low
temperature \cite{07}, only negligible changes, i.e., less than
$0.6 \%$, were observed. Thus, attempts at substantial
modifications of the nuclear lifetime have failed.

Nevertheless, it is a very attractive problem to modify the
lifetime, equivalently the energy level width, because the nuclear
waste problem must be solved. If the lifetime becomes shorter,
waste cleaning processes may be accelerated. While, prolongation
of the lifetime implies suppression of radioactivity. Furthermore,
narrowing the width (equivalently lengthening the lifetime) would
be useful for longer storage of radioactive material for some
purposes. A more important point is that precision measurement of
the $\gamma$-ray spectra may be improved. Namely, the accuracy of
the M{\" o}ssbauer experiment might be improved if the width of
its absorption spectrum could made narrower.

On the other hand, it is also known that spatial structure of the
vacuum field can change the atomic and nuclear energy levels and
widths \cite{08, 09}. That is to say, if space is limited by two
perfect conducting plates on the surface of which all wave
functions vanish, the vacuum field becomes discrete and,
therefore, induces some modification of physical quantities.
However, observable effects could only be obtained for plates with
separations on the order of micrometers \cite{08, 09, 10, 11, 12,
13, 14}.

A M{\"o}ssbauer experiment carried out to observe nuclear energy
level shifts \cite{11,12,13} discovered that the width became
narrower \cite{15}; then, the data were carefully reanalyzed
\cite{16}. Usually, broadening occurs easily due to various noises
but narrowing is very difficult. Therefore, it is very interesting
to investigate the mechanism of such a phenomenon which must be
different from the effects of the chemical environment.  We should
stress here that we are not talking about a reduction of
broadening due to various noises but a reduction of the natural
line width itself.

 A novel idea has been proposed \cite{16}  to
 explain the phenomenon found in the M{\"o}ssbaur experiment.
 If the photon emitted from the source could partially return and be reabsorbed
 by the original source, the duration for the source nucleus to stay in the
 excited state would effectively increase; and consequently,
 the nuclear half-life could be prolonged, equivalently, the width
 of this state  could become narrower.  In free space, return
 of the emitted photon to the source is impossible. However,
 the photon may be forced to return to the source by operating
 reflectors, say metallic plates or a cylinder. A process in which
 even a part of the photon returns and is
 reabsorbed by the source nucleus after backscattering on the
 metal surface would cause suppression of photon emission. When such a process
 is repeated many times, the lifetime is finally prolonged. Of
 course, all these processes such as emission, backscattering and
 absorption should occur elastically, i.e., without any
 energy loss. Therefore, the source nucleus should be implanted in
 a solid.
 One step of photon reabsorption would cause only a tiny modification
 of the lifetime, but iterative processes would proceed
 step by step until the probability for finding the nucleus in the
 excited state would become half \cite{17,18,19,20,21}.

\section{Description of the process}

\subsection{Decay Equation and the Half-life}

Let $|\psi_0 (t)|^2$ be the probability for finding the system in
the state $\psi_0 (t)$ at the time $t$, whether a single-particle
or many-particle system. This state is assumed to be unstable and,
therefore, decays, i.e., by $\gamma$ emission in our case. Then,
its decay equation reads
\begin{equation}
{d \over dt} |\psi_0 (t)|^2 = - \lambda |\psi_0 (t)|^2 ,
\end{equation}
where $\lambda $ is the decay constant. Thus, the state can
generally be expressed as
\begin{equation}
\psi_0 (t) = A \: exp \:[-  {i \over \hbar } (E_0 - {i \Gamma
\over 2} )t] .
\end{equation}
The state has a complex energy eigenvalue because it is unstable,
and $\Gamma$ is the width of the state. The state $\psi_0 (t)$
given in Eq. (2) satisfies, of course, Eq. (1), and $\lambda =
\Gamma /\hbar$.

The validity of the expression in Eq. (2) for $\psi_0 (t) $ can
immediately be verified by taking a Fourier transform of $\psi_0
(t) $, i.e.,
\begin{equation}
\phi (E) = {A \over \sqrt{2 \pi}} \:  {i \hbar \over [(E-E_0 ) + i
{\Gamma \over 2 } ]} ;
\end{equation}
then,
\begin{equation}
|\phi (E)|^2 = {A^2 \over {2 \pi}} \:  { \hbar^2 \over [(E-E_0 )^2
+ ({\Gamma \over 2 })^2 ]}.
\end{equation}
This last equation represents a Lorentzian spectrum, and $\Gamma$
is definitely the width of the state $\psi_0 (t)$.

Since the decay rate cannot be  measured experimentally with a
single-particle system, the measurement is always carried out by
means of a particle assembly. However, one can obtain an identical
answer by repeating the measurements again and again with a single
particle under the same conditions. This is due to the fundamental
concept of quantum mechanics. Therefore, we investigate the
process with an assembly of nuclei instead of a single nucleus,
which is fundamentally the same as phenomena occurring with a
single nucleus.

If emitted photons return once to the source after being scattered
coherently by a metal surface and are reabsorbed, the decay
equation with the decay constant $\lambda$ is given as
\begin{equation}
  dN = -  {\lambda N }  \: dt + \Sigma\: \lambda N \:dt \equiv - \lambda^{(1)} N\: dt,
\end{equation}
where $ \lambda^{(1)} = (1- \Sigma ) \lambda $. Even if $N$ is
replaced by $|\psi_0 (t)|^2$, this equation holds as it is.
 $\Sigma$ denotes the probability associated with
photon reabsorption. Therefore, the second term stands for the
effect of $\gamma$ reabsorption. Equation (5) is valid for the
time interval $t_0 \leq t < 2t_0 $ ($t_0 = 2R/c$, $c=$speed of
light, and $R$ is the distance between the source and the metallic
surface where the photon is scattered) during which the photon
returns only once. Thereby, the decay constant $\lambda$ is
effectively modified as $\lambda^{(1)} $.

For the $m$th return of the photon, i.e., for the time interval
$mt_0 \leq t < (m+1)t_0$, the decay equation is generally
expressed as
\begin{equation}
dN_{m+1} = - \lambda^{(m)} \:  N_{m+1} \: dt \; .
\end{equation}
Integration of this equation over that time interval yields
\begin{equation}
N_{m+1} = N_m exp[- \lambda^{(m)} \: (t_0 - \epsilon)] \; ,
\end{equation}
where the limit $\epsilon \rightarrow 0 $ should be taken in the
final stage and
\begin{equation}
\lambda^{(m)} =  (1-\Sigma)^m \: \lambda \:.
\end{equation}
For $0 \leq t <  t_0 $, the photon has no time to return to the
source, i.e., $m=0$; therefore, $N_1 = N_0 \: exp[-\lambda \:(t_0
- \epsilon )]$, where $\lambda = \lambda^{(0)}$. By iteration, we
find
\begin{eqnarray}
N_{m+1} &=& N_0 \prod_{s=0}^m exp\:(-\lambda^{(s)} t_0 ) \nonumber\\
        &=& N_0 \:exp\:[- \lambda {{1-(1-\Sigma)^{m+1} } \over
        \Sigma}\: t_0 \:]
\end{eqnarray}
in the limit $\epsilon \rightarrow 0 $. For $m \rightarrow 0$, we
have $N_1 = N_0 exp(- \lambda t_0 )$. Similarly, for $\Sigma
\rightarrow 0 $, it becomes $N_{m+1} = N_0 \: exp[-\lambda (m+1)
t_0 ] = N_0 \: exp(-\lambda t )$, where $(m+1) t_0 = t$. This
result is usual; i.e., nothing changes because of $\Sigma =0$.
Equation (9) indicates that the decay constant is changed step by
step at every stage in the reabsorption of returning photons. Let
us find the value of $m$ by setting $N_{m+1} = {1 \over 2} N_0 $
because the number of radioactive nuclei becomes half of the
initial amount at the $(m+1)$th step. Namely,
\begin{equation}
    ln 2 = \lambda \: {1- (1-\Sigma)^{m+1} \over \Sigma } \: t_0 ,
\end{equation}
which gives
\begin{equation}
     m+1 = { ln [1- ( \tau_{1/2} / t_0 ) \Sigma ] \over ln (1-\Sigma ) }
         = { ln [1- ( c \: \tau_{1/2} / 2R ) \Sigma ] \over ln (1-\Sigma )
         },
\end{equation}
where we used $(ln2)/\lambda = \tau_{1/2} $ and $t_0  = 2R/c $. By
introducing the effective decay constant ${\tilde \lambda}$ in Eq.
(9) for $ \Sigma \neq 0$, it can be rewritten as
\begin{equation}
     N_{m+1} = N_0 \: exp \: (- {\tilde \lambda t)}
\end{equation}
with
\begin{equation}
     {\tilde \lambda } = \lambda \: { 1- (1- \Sigma )^{m+1} \over
     (m+1)\: \Sigma } \: .
\end{equation}
Rewriting Eq. (11) as
\begin{equation}
 (1-\Sigma)^{m+1} =1-\left ( {c \tau_{1/2} \over 2R}\right)  \Sigma
\end{equation}
and substituting this result with Eq. (11) into Eq. (13), we
obtain the expression of ${\tilde \lambda}$ as
\begin{equation}
{\tilde \lambda} = \lambda \left ( {c \tau_{1/2} \over 2R}\right)
{ ln [1- ( c \: \tau_{1/2} / 2R ) \Sigma ] \over ln (1-\Sigma )}
\: .
\end{equation}
As we explained in the beginning, the discussion here with an
assembly of $N$ nuclei is actually independent of $N$, and the
same result can be obtained with the state $\psi_0 (t) $. Namely,
${\tilde \lambda}$ is the modified decay constant of the state,
and ${\tilde \Gamma} = \hbar {\tilde \lambda}$ is the modified
width of the state.

 Because $0<{\tilde \lambda }/ \lambda <1$, the decay is
delayed; thus, the level width appears narrower. Accordingly, the
modified lifetime is now found to be
\begin{equation}
     {\tilde \tau}_{1/2} = ({2R \over c }) \:
        { {ln [1- ( c \: \tau_{1/2} / 2R ) \Sigma ]} \over {ln (1-\Sigma )}
        }.
\end{equation}
 As we have seen above, the photon reabsorption process is
repeated m times before the half-life of the state is formed.
Indeed, that process directly participates to build the half-life.
It is not simple radiation trapping, but in the course of photon
reabsorption, the nuclear lifetime has been gradually build up
step by step.

\subsection{The Backscattering Cross Section}

Let us now investigate the $\gamma -$backscattering by the
metallic cylinder. The elastic scattering of $\gamma-$rays from
the metal surface is coherent and mostly caused by atomic
electrons. For an incident photon energy much larger than the
atomic binding energy, the scattering can be described in a good
approximation by the seagull term of the corresponding Feynman
diagrams, i.e., by
\begin{equation}
{d \sigma  \over d \Omega} = \sum_{\alpha,\: \alpha^{\prime}}
 ({{e^2} \over {m_e c^2}} )^2 \: |\:\epsilon^{(\alpha)} \cdot
 \epsilon^{(\alpha^{\prime})} |^2  \:|F(E_\gamma,\theta)|^2,
\end{equation}
where $\epsilon^{(\alpha)}$ $(\alpha=1,2)$ is the photon
polarization vector, $m_e$ and $e$ are the electron mass and
charge, respectively, $c$ is the velocity of light, and
$F(E_{\gamma} , \theta)$ is the form factor given by the $\gamma
-$ray energy $E_{\gamma}$ and scattering angle $\theta $. With the
Cartesian components of $\epsilon^{(\alpha)}=(1,0,0) $ for $\alpha
=1,2$ and
\begin{eqnarray}
\epsilon^{(\alpha^{\prime})} = \left\{
\begin{array}{l}
(sin \phi, - cos \phi, 0)  ~~~~~~~~~~~~~~~~~~(\alpha^{\prime} =1)\; , \\
(cos\theta \: cos\phi, cos\theta \: sin \phi, -sin\theta)
~~(\alpha^{\prime} =2) \; ,
\end{array}\right.
\end{eqnarray}
we find
\begin{equation}
\sum_{\alpha,\: \alpha^{\prime}} |\:\epsilon^{(\alpha)} \cdot
 \epsilon^{(\alpha^{\prime})} |^2 = sin^2 \phi + cos^2 \theta \: cos^2
 \phi  \; .
\end{equation}
For unpolarized beams, the differential cross-section appears in
the form
\begin{eqnarray}
{d \sigma  \over d \Omega}
 &=& {1 \over 2} \left[ {d \sigma  \over d \Omega} \: (\phi =0 )
   +  {d \sigma  \over d \Omega} \: (\phi = {\pi \over 2 } ) \right] \\ \nonumber
 &=& ({{e^2} \over {m_e c^2}} )^2 \: |F(E_\gamma,\theta)|^2 \:
  {1 \over 2} (1+cos^2\theta) \: .
\end{eqnarray}
 Since photon coherent scattering by protons in the nucleus
may take place simultaneously, one has to take it into account.
This scattering can be described in analogy with the atomic case,
provided the electron mass is replaced by the proton mass.
However, its contribution is actually negligible compared to that
of the scattering by atomic electrons because the proton mass is
much larger than the electron mass.

Now, only photons scattered entirely backward can successfully
return to the source to be reabsorbed. Assuming the nucleus to be
a point particle because the nucleus is much smaller than the
photon wavelength, one may express the backward scattering cross
section $\sigma_{\pi} $ as
\begin{equation}
\sigma_{\pi} = \int {d \sigma  \over d \Omega} \: \delta(cos
\theta - cos \pi ) \: \delta(\phi - \phi_0 )\: d\Omega = ( {{e^2}
\over {
 m_e  c^2 } })^2  |F(E_\gamma , \pi)|^2 .
\end{equation}

Since the photon is scattered by the atoms in a metallic cylinder
of thickness $d$, we must count the number of atoms per $cm^2$.
This number can be given by $nd$, where $n$ is the number of atoms
per $cm^3 $, and can be obtained from the density divided by the
atomic mass: $n= \rho /M = \rho N_A /A $ with Avogadro's number
$N_A$. When the $\gamma - $ray comes from a direction at a angle
of $\psi$ from the normal direction of the cylinder surface, $d$
must be replaced by $d_1 = d/ cos \psi$. For this case, the number
of atoms per $cm^2 $ should be $n_1 =nd_1 = nd/cos \psi$.

Since the scatterers are bounded in the solid, the effect of
lattice vibration should be taken into account. It can be done by
introducing the Debye-Waller factor \cite{21}. Therefore, backward
scattering cross section $\sigma_\pi $ should be multiplied by
this factor .

\subsection{Expression of $\Sigma$}

Generally, radioactive nuclei emit $\gamma-$rays isotropically, so
the total number of photons emitted during the time $dt$ is given
by
\begin{equation}
\int \rho_u \: dS_u = \lambda N dt \;,
\end{equation}
where $\lambda$ is the decay constant, $N$ is the number of
radioactive nuclei at a certain time, $dS_u $ is an element of
area on a sphere of arbitrary radius $u$, and $\rho_u$ is the
surface density of photons passing through this area, when the
initial number of photons is $\lambda N dt$. $\rho_u$ is given by
\begin{equation}
\rho_u = { \lambda N dt \over 4\pi u^2} \;.
\end{equation}
This relation holds for the sphere of any arbitrary radius, i.e.
$\rho_R = { \lambda N dt / (4\pi R^2 )} \;.$

Let us consider a element of area $dS_z $ at a point on the
cylinder surface that is located at the distance $u$ from the
center. Since $z=R \: tan\psi $, we find
\begin{eqnarray}
dS_z &=&  dz  \: R \: d\theta = { R^2 \over cos^2 \psi } \: d\psi
                   \: d\theta       \nonumber  \\
     &=& ({R \over u} )^2 { 1 \over sin\psi \: cos^2 \psi } \:
         ( u^2 sin\psi \: d \psi \: d\theta )  \nonumber \\
     &=& {1 \over sin\psi } \: dS_u \; ,
\end{eqnarray}
where the relations $R =u\: cos\psi$ and $dS_u = u^2 \: sin\psi \:
d\psi \: d\theta $ are used. Then, we obtain
\begin{eqnarray}
\rho_u \: dS_u
  &=& ({ \lambda N dt \over 4\pi u^2 }) \: sin\psi \: dS_z
   = ({ \lambda N dt \over 4\pi R^2 }) \: ({R \over u})^2 \: sin\psi \:
     dS_z \nonumber  \\
  &=&({ \lambda N dt \over 4\pi R^2 })
                               \: cos^2 \psi \: sin\psi \: dS_z
   \equiv \rho_z (\psi) \: dS_z \; ,
\end{eqnarray}
where
\begin{equation}
\rho_z (\psi) = \left( {cos^2 \psi \: sin\psi \over 4 \pi R^2 }
\right) \lambda N dt \equiv {\hat \rho_z }(\psi) \lambda N dt
\end{equation}
is the surface density of photons on an element of area $dS_z$ of
the cylinder when an initial number of photons, $\lambda N dt$,
comes. ${\hat \rho}_z $ is the surface density of photons on
$dS_z$ when a single incident photon comes in. Of course, the
total number of photons emitted during $dt$ can be obtained by
integrating over a cylinder surface of infinite length
\begin{equation}
\int \rho_z (\psi)\: dS_z = 2 \int_0^{\pi /2} \left( {\lambda N dt
\over 4 \pi R^2 } \: cos^2 \psi \: sin\psi \right) {2 \pi R^2
\over cos^2 \psi }\: d\psi =  \lambda N dt \; .
\end{equation}
In addition, a photon that needs to travel a distance $u$
contributes $(R/u)$ times as much as one travelling $R$.

Thus, $\Sigma$ is expressed by an integral over the cylinder
surface; i.e.,
\begin{eqnarray}
\Sigma &=& \zeta \int ( {R \over u} \: n_1 f\: \sigma_\pi )\:
 {\hat \rho}_z \: dS_z \nonumber \\
 &=& \zeta  \int [ cos \psi ({nd \over cos
\psi } )\: f\: \sigma_{\pi} ] \left( {cos^2 \psi sin \psi \over
4\pi R^2 } \right) {R^2 \over cos^2 \psi } \: d \psi \: d\theta\\
\nonumber
 &=& 2 \zeta \: {2\pi \over 4 \pi} \int_0^{\psi_L} (nd \: f \sigma_\pi
 )\: sin \psi \: d\psi \\ \nonumber
 &=& \zeta \:  n d \: f \:\sigma_\pi
     \left[1 - \{ 1+ ({L_0 \over 2R})^2 \}^{-1/2} \right] \;,
\end{eqnarray}
where $f$ is the Debye-Waller factor \cite{21} and $\zeta$ is the
photon absorption probability of the source. $R$ and $L_0$ are the
radius and the length of the cylinder, respectively; therefore,
$\psi_L= arc tg(L_0/2R)$.

\subsection{Photon Absorption Probability}

 The photon absorption probability may be given by the ratio of the
 $\gamma -$absorption cross section to the total cross section:
\begin{equation}
\zeta = { \sigma_{\gamma}^N \: f_1 \over \sigma_{tot} } \; ,
\end{equation}
where
\begin{equation}
\sigma_{tot} = f_1  \:(\sigma_{\gamma}^N  + \sigma_{coh}^N ) +
\sigma_{pe}^N  + \sigma_{incoh}^N  +  f_1 \: \sigma_{coh}^A +
\sigma_{pe}^A  + \sigma_{incoh}^A \; .
\end{equation}
The superscripts $N$ and $A$ denote nuclear and atomic processes,
respectively. $f_1 $ is again the Debye-Waller factor for the
source nuclei. $\sigma_{\gamma}^N $ is the $\gamma$-absorption
cross section given as \cite{22}
\begin{equation}
{\sigma_{\gamma}^N } = 2 \pi \; {{\bar \lambda_{\gamma} } }^{2} \;
\frac {2 J_f + 1} {2J_i +1} \;  \frac {1} {1+\alpha } \; ,
\end{equation}
where ${ \bar \lambda_{\gamma} } = \hbar c /E_\gamma $ and $\alpha
$ is the internal conversion coefficient. $J_i $ and $J_f $ are
the spins of the initial and the final nuclear states,
respectively. The nuclear photoelectric absorption cross section
$\sigma_{pe}^N $ can be obtained from the relation $\sigma_{pe}^N
= \sigma_{\gamma}^N \alpha $. The cross sections of coherent and
incoherent scattering of gamma rays from a nucleus are calculated
by the formulae given in Ref. 23. They are negligible small
compared with $\sigma_{pe}^N $ and $ \sigma_{\gamma}^N$. The cross
sections of coherent and incoherent scattering and of
photoelectric absorption of photons by atoms are also given in
Ref. 23.

The validity of Eq. (29) was already examined for the $CsCl$
compound \cite{21}. All necessary cross sections were obtained
from the XCOM Photon Cross Section Database \cite{24} and $\alpha
=1.72$ \cite{25}. The result was
\begin{equation}
\zeta_{CsCl} = 8.3 \times 10^{-3}
\end{equation}
at $T=4.2 \: K$, i.e., the probability of $\gamma-$absorption by
$Cs$ in the $CsCl $ compound is about $0.8 \: \%$. This value can
be compared to the relative depths of the absorption spectra
observed in M{\"o}ssbauer experiments, which are all about $0.7
\sim 4 \%$ \cite{16, 26, 27, 28, 29, 30, 31, 32}. The agreement is
good, so the value of $\zeta$ calculated using Eq.(29) is
reliable.

\section{Numerical Calculation}

Let us examine our theory. As above, the conditions to maximize
the effect is to select a material which has a large Debye
temperature and which induces large backward scattering.
Furthermore, the energy of emitted gamma-ray, $E_\gamma $, should
be less than $100 \: keV$. Otherwise, the Debye-Waller factor
becomes very small, and the effect is greatly reduced. If
$E_\gamma$ is less than $10 \: keV$, various noises associated
with detectors become large, and clean data may not be obtained.

\subsection{Gamma-ray Source}

Considering the above conditions, we try to examine the first
excited state, the ${5 \over 2 }^+ $ state of ${}^{133} Cs $,
which is $81$-keV level with a lifetime of $6.27 \: ns $. To
eliminate the recoil effect, this nucleus should be implanted in a
solid. When a compound ${}^{133} BaTiO_3 $ is taken, ${}^{133} Ba
$ decays into ${}^{133} Cs $ through the electron conversion
process because ${}^{133} Ba $ is radioactive, so ${}^{133}
Cs^\ast $ remains in the compound. Of course, ${}^{133} Cs^{\ast}
$ is in the first excited  ${5 \over 2 }^+ $ state and emits a
$81$-keV gamma ray when it drops into the ground  ${7 \over 2 }^+
$ state.

The compound ${}^{133} BaTiO_3 $ has a perovskite structure with a
rather high Debye temperature, $\theta_D = 431.8 \: K$ \cite{31}.
Although, the Debye temperature of the perovskite resulting from
the decay of ${}^{133} Ba $ into ${}^{133} Cs$ is not actually
known, it may be assumed to be the same as that of ${}^{133}
BaTiO_3 $ because both of them have the perovskite structure and
${}^{133} Ba $ simply converts to ${}^{133} Cs$ through the EC
process. Therefore, the Debye temperature $\theta_D = 431.8 \: K $
is taken for $ Cs_2 Ti O_3 $. Then, the Debye-Waller factors can
be found as $f_1 = 0.3434$, $0.3407$, and $0.2750$ at temperatures
$T=4.2 \: K$, $15 \: K $, and $77 \: K $, respectively.

All the cross sections necessary to estimate the gamma absorption
probability are obtained using Eq. (31) and the relations
$\sigma_{pe}^N = \alpha\sigma_{\gamma}^N $ and $\sigma_{incoh}^N =
(A/Z)^2 \sigma_{coh}^N$, where $A$ and $Z$ denote the nuclear mass
number and the atomic number, respectively. $\sigma_{coh}^N$ has
been estimated with Eq. (20), where the electron mass was replaced
by the the proton mass and with the fact that the normalized
nuclear form factor is almost unity in the energy region
considered here.

 The cross sections for the atomic processes should be
calculated with the $Cs_2TiO_3$ compound. They can be obtained by
using the XCOM Photon Cross Section Database with the modified
relativistic form factor \cite{24}. The results are listed in
Table 1. Thus, the absorption probabilities are $\zeta =0.166$,
$0.165$, and $0.138$ at $T=4.2 \: K$, $15 \: K $, and $77 \: K $,
respectively.

\subsection{Materials of the cylinder}

Platinum is suitable for a large backward scattering of
$\gamma-$rays because the atomic number of platinum is $Z=78$ and
its Debye temperature is $\theta_D = 240 \: K$. Therefore, the
Debye-Waller factors for $E_\gamma = 81 \: keV$ are $f =0.271$,
$0.263$, and $0.125$ at $T=4.2 \: K$, $15 \: K $, and $77 \: K $,
respectively.

The modified relativistic form factor for $Z=78$ is estimated,
using the XCOM program \cite{24}, as $F(81 \: keV, \pi ) = 3.3442$
at $\theta = \pi$. Accordingly, the cross section of the $81$-keV
gamma-ray backward scattered by platinum is $\sigma_\pi =0.888
\times 10^{-24} \: cm^2 $. The validity of the XCOM program has
been verified by experiments \cite{01, 02}. Since the density of
platinum is $\rho = 21.41 \: (g/cm^3 )$, the number of atoms per
$cm^2 $ is $n=6.58 \times 10^{22} \: cm^{-2}$. A platinum cylinder
with a thickness of $d=0.05 \: cm$, inner diameters of
$2R=0.2,~0.3, ~0.5$, and $1.0 \: cm$, and a length of $L_0=5\: cm$
is used.

\subsection{Results for the Energy Level Width and Lifetime}

With all the information obtained above, the level width and
lifetime can be calculated using eqs. (15) and (16). The results
are given in Table 2. They are within the measurable range. The
value of $m$ in Eq. (15) depends on both the temperature and the
cylinder radius $R$ and ranges between $190$ and $650$.

Gold is also a suitable material as a photon reflector. Its Debye
temperature is $165 \: K$, which gives the Debye-Waller factors of
$0.150$, $0.137$, and $0.0102$ at $T=4.2\: K$, $15\:K$, and
$77\:K$, respectively, for $E_{\gamma}=81\:keV$. Since the
modified relativistic form factor is $F(81 \:keV,
\theta=\pi)=3.4057$, the total cross section at $\theta=\pi$ is
$\sigma_\pi = 0.921 \times 10^{-24} \: cm^2$. The density of gold
is $18.85 \:(g/cm^{-3} )$ and, so $n=5.76\times 10^{22}
\:cm^{-2}$. The results for this case are shown in parentheses of
in Table 2.

Notice that ${\tilde \Gamma} \rightarrow \Gamma$ and $ {\tilde
\tau}_{1/2} \rightarrow \tau_{1/2}$ in the limit of $R \rightarrow
\infty $. Their temperature dependence appears through the
Debye-Waller factor of the metallic cylinder. At room temperature,
the width and the lifetime do not change.

\section{Conclusion}

As is seen above, the decay constant $\lambda$ is altered step by
step at every stage of $\gamma$ reabsorption. $\lambda$ is
actually related to the level width as $\Gamma = \hbar \lambda $.
Therefore, the process definitely changes the level width, and
equivalently the half-life. Our results imply that the accuracy in
M{\" o}ssbauer measurements can be improved by setting both the
source and the absorber, respectively, between two plates.

The spatial structure of vacuum field can also change the atomic
and the nuclear lifetimes \cite{14, 34}. However, in that case,
plausible effects appear only for much smaller separations between
the two plates. Therefore, it is negligible in the present
investigation.

Other processes, so-called "radiation trapping", were also
investigated \cite{35, 36}, and a prolonged nuclear lifetime was
observed \cite{37}. The interpretation was that the time evolution
of gamma-ray emission was modulated as a result of the time
consumed during photon exchange between two radioactive nuclei.
Namely, the photon is delayed in coming out of the system while
the two nuclei play with the photon. However, it has no relevance
to the lifetime unless the energy level width is modified. The
level width and lifetime can be changed only when population of
the excited state is increased through the mechanism discussed
above. In conclusion, a sharper spectrum can be obtained by using
the method proposed here, and the accuracy in measuring the
$\gamma-$ray spectrum can be improved. Radiation trapping can
occur even without modification of energy levels, but such a
simple "trapping" cannot have any effect on the M{\"o}ssbauer
spectrum connected directly to the energy level width.
Measurements should be carried out at low temperatures with
specific detectors with good timing performance and good energy
resolution.

\newpage
\begin{figure}
\begin{center}
\includegraphics[width=5cm, height=6cm]{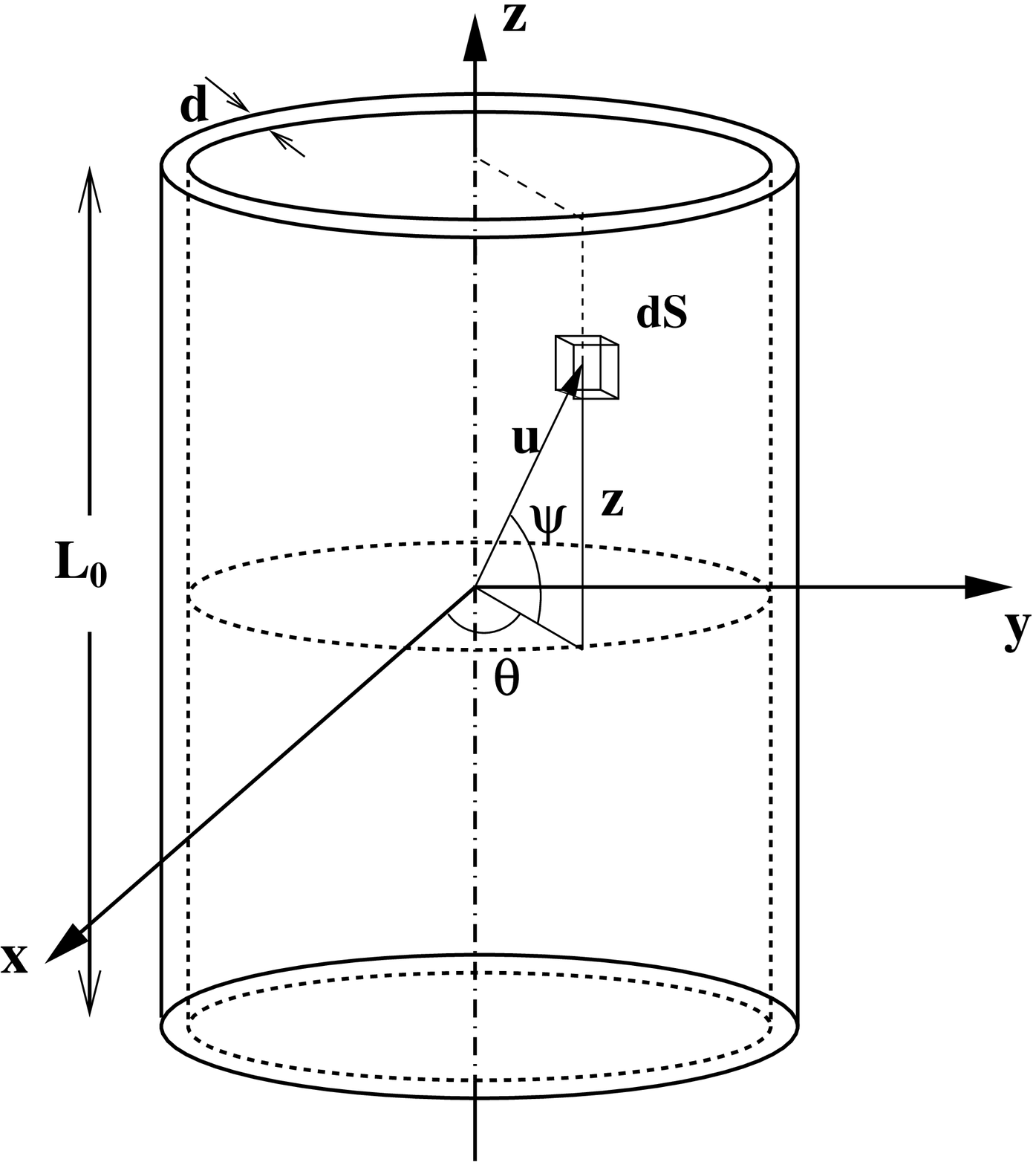}
\end{center}
\caption{Geometry of a cylinder.} \label{fig.1}
\end{figure}

\newpage
\begin{figure}[tb]
\begin{center}
\includegraphics[width=2.5in, height=2.0in]{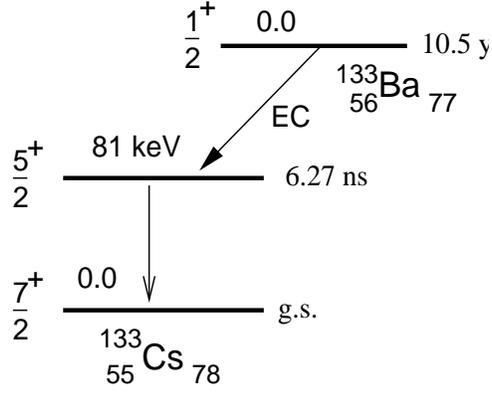}
\end{center}
\caption{Decay scheme.} \label{fig.2}
\end{figure}

\newpage
\begin{table}[tb]
\caption{Cross sections in unit of $10^{-19} cm^2$. Superscripts
$N$ and $A$ denote the nuclear and atomic processes,
respectively.} \label{table1}
\begin{center}
\begin{tabular}{@{\hspace{\tabcolsep}\extracolsep{\fill}}rcccc}
\hline\hline
     Nuclear process \vline & $\sigma_{\gamma}^N$ &  $\sigma_{pe}^N$ & $\sigma_{coh}^N$
      & $\sigma_{incoh}^N$   \\\cline{2-5}

              \vline& $1.03$ & $1.77$ &  $9.99\times 10^{-10}$ & $~5.84\times 10^{-9}$
             \\
\hline\hline
      Atomic  process \vline & $\sigma_{pe}^A$ &  $\sigma_{coh}^A$ & $\sigma_{incoh}^A$
      &    \\\cline{2-5}
     \vline& $~5.60\times 10^{-3}$ & $~2.69\times 10^{-4}$ &
     $~1.98\times 10^{-4}$ &              \\
\hline\hline
\end{tabular}
\end{center}
\end{table}

\newpage
\begin{table}[tb]
\caption{Modified widths and lifetimes. Values are obtained with a
platinum cylinder, and those in the parentheses are with a gold
cylinder.  The standard value is $ \tau_{1/2} = (6.27 \pm 0.02) \:
(ns)$ and $\Gamma = 7.28 \times 10^{-8} eV$. $ \Delta \Gamma =
{\tilde \Gamma} - \Gamma$ and $ \Delta \tau = {\tilde \tau}_{1/2}
- \tau_{1/2} $ .} \label{table2}
\begin{center}
\begin{tabular}{@{\hspace{\tabcolsep}\extracolsep{\fill}}ccccccc}
\hline \hline
     $T(K)$ & $R(cm)$ &  $\Sigma (10^{-4})$ & ${\tilde \Gamma} (10^{-8}
     eV)$  & $\Delta \Gamma / \Gamma(\%)$  & $ {\tilde \tau_{1/2}} \: (ns)$
     & $ {\Delta \tau} /{\tau_{1/2} } \:(\%)$ \\
\hline
     4.2    & 0.10 &  1.280(0.6457)     &          6.83(7.05)     & -6.14(-3.07)
             &  6.68(6.47)     &          +6.54(+3.16)   \\
            & 0.15 &  1.228(0.6194)     &          6.99(7.13)     & -3.90(-1.95)
             &  6.52(6.39)     &          +4.05(+1.99)  \\
            & 0.25 &  1.176(0.5933)     &          7.11(7.19)     & -2.23(-1.12)
             &  6.41(6.34)     &          +2.27(+1.13)   \\
            & 0.50 &  1.050(0.5297)     &          7.20(7.24)     & -0.986(-0.496)
             &  6.33(6.30)     &         +0.995(+0.499)   \\
\hline
     15    & 0.10 &  1.233(0.5836)     &          6.85(7.07)     & -5.91(-2.77)
             &  6.66(6.45)     &          +6.28(+2.85)   \\
            & 0.15 &  1.183(0.5598)     &         7.00(7.15)     & -3.75(-1.76)
             &  6.51(6.38)     &          +3.90(+1.79)  \\
            & 0.25 &  1.133(0.5362)     &          7.12(7.20)     & -2.14(-1.01)
             &  6.41(6.33)     &          +2.19(+1.02)   \\
            & 0.50 &  1.011(0.4787)     &          7.21(7.24)     & -0.949(-0.449)
             &  6.33(6.30)     &         +0.958(+0.451)   \\
\hline
     77    & 0.10 &  0.4704(0.06621)     &          7.11(7.25)     & -2.22(-0.311)
             &  6.41(6.29)     &          +2.28(+0.312)   \\
            & 0.15 &  0.4513(0.06351)    &          7.17(7.26)     & -1.42(-0.199)
             &  6.36(6.28)     &          +1.44(+0.199)  \\
            & 0.25 &  0.4322(0.06084)     &          7.22(7.27)     & -0.813(-0.114)
             &  6.32(6.28)     &          +0.820(+0.114)   \\
            & 0.50 &  0.3859(0.05431)    &          7.25(7.27)     & -0.361(-0.0508)
             &  6.29(6.27)     &         +0.363(+0.0508)   \\
\hline \hline
\end{tabular}
\end{center}
\end{table}


\newpage

\begin{thebibliography}{99}
\bibitem{01} K. Sidappa, N. Govinda Nayak, K. M. Balakrishna, N.
Lingappa, and Shivaramu, Phys. Rev. \textbf{A39}, 5106 (1989).
\bibitem{02} I. S. Elyaseery, A. Shukri, C. S. Chong, A. A. Tajuddin, and
             D. A. Bradley, Phys. Rev. \textbf{57}, 3469 (1998).
\bibitem{03} K. T. Bainbridge, M. Goldhaber, and E. Wilson, Phys. Rev. \textbf{
90}, 430 (1953).
\bibitem{04} H. Mazaki and S. Shimizu, Phys. Rev. \textbf{148}, 1161 (1966).
\bibitem{05} T. Ohtsuki, H. Yuki, M. Muto, J. Kasagi, and K. Ohno,
Phys. Rev. Lett. \textbf{93}, 112501-1 (2004).
\bibitem{06} K. T. Bainbridge, Chem. Eng. News \textbf{30}, 654 (1952).
\bibitem{07} D. H. Byers and R. Stump, Phys. Rev. \textbf{112}, 77 (1958).
\bibitem{08} Il-T. Cheon, Phys. Rev. \textbf{A37}, 2785 (1988).
\bibitem{09} Il-T. Cheon, Laser Physics \textbf{4}, 579 (1994).
\bibitem{10} Il-T. Cheon and S. D. Oh, J. Phys. Soc. Jpn \textbf{58}, 1581 (1989).
\bibitem{11} Il-T. Cheon, Hyperfine Interactions \textbf{78}, 231 (1993).
\bibitem{12} Il-T. Cheon, J. Phys. Soc. Jpn \textbf{63}, 47 (1994).
\bibitem{13} Il-T. Cheon, J. Phys. Soc. Jpn \textbf{63}, 2453 (1994).
\bibitem{14} Il-T. Cheon, Z. Phys. \textbf{D39}, 3 (1997).
\bibitem{15} Il-T. Cheon, presented at the 1995 SNU workshop on
 "Application of Field Theory"(Seoul), and at the 1999 Autum Meeting of
 Phys. Soc. Jpn; "Genshikaku Kenkyu" \textbf{44}, 5 (1999).
\bibitem{16} Il-T. Cheon, J. Phys. Soc. Jpn \textbf{70}, 3193 (2001).
\bibitem{17} Il-T. Cheon and M. T. Jeong, J. Korean Phys. Soc. \textbf{43}, S87 (2003).
\bibitem{18} M. T. Jeong, Hyperfine Interactions \textbf{156/157}, 165 (2004).
\bibitem{19} Il-T. Cheon and M. T. Jeong, J. Korean Phys. Soc. \textbf{46}, 397 (2005).
\bibitem{20} Il-T. Cheon and M. T. Jeong, J. Korean Phys. Soc. \textbf{47}, 944 (2005).
\bibitem{21} Il-T. Cheon, Europhys. Lett. \textbf{68}, 900 (2005).
\bibitem{22} E. U. Condon and H. Odishaw: \textit{Handbook of
Physics 2nd edition}, (McGraw-Hill, New York, 1967) pp. 9-190.
\bibitem{23} Janos Kirz: \textit{X-ray Data Booklet} (Lawrence Berkley National
Laboratory, 2000), Sec. 3.1.
\bibitem{24} J. W. Strutt, XCOM Photon Cross Section Database,
http://www.cxro.lbl.gov.
\bibitem{25} Yu. V. Sergeenkov and V. M. Sigalov, Nucl. Data Sheets for A=133
\textbf{49}, 639 (1986). This value was revised later.
\bibitem{26} A. J. F. Boyle and G. J. Perlow, Phys. Rev. \textbf{149}, 165 (1966).
\bibitem{27} L. E. Campbell and G. J. Perlow, Nucl. Phys. \textbf{A109}, 59 (1968).
\bibitem{28} L. E. Campbell, G. L. Montet, and G. J. Perlow, Phys. Rev. \textbf{B15}, 3318 (1977).
\bibitem{29} E. Verbiest, H. Pattyn, and J. Odeurs, Nucl. Instr. Methods \textbf{182/183}, 515 (1981).
\bibitem{30} H. Pattyn, J. Odeurs, and E. Verbiest, Nucl. Instr. Methods \textbf{170}, 399 (1980).
\bibitem{31} G. Langouche, Hyperfine Interactions \textbf{29}, 1283 (1986).
\bibitem{32} H. Muramatsu et al., Phys. Rev. \textbf{B58}, 11313 (1998).
\bibitem{33} W. N. Lawless, Phys. Rev. \textbf{B17}, 1458 (1978).
\bibitem{34} V. I. Vzsotskii, Phys. Rev. \textbf{C58}, 337 (1998).
\bibitem{35} T. Holstein, Phys. Rev. \textbf{72}, 1212 (1947).
\bibitem{36} J. Huennekens, H. P. Park, T. Colbert, and S. C.
McClain, Phys. Rev. \textbf{A35}, 2892 (1987).
\bibitem{37} A. J. Chumakov, J. Metge, A. Q. R. Baron, R. B{\"u}ffer,
Yu. V. Shvyd'ko, H. Gr{\"u}nsteudal, and H. F. Gr{\"u}nsteudal,
Phys. Rev. \textbf{B56}, R8455 (1997).
\end{thebibliography}
\end{document}